# Image sensor array based on graphene-CMOS integration

**AUTHORS:**

STIJN GOOSSENS[1,†], GABRIELE NAVICKAITE[1,†], CARLES MONASTERIO[1,†],

SHUCHI GUPTA[1,†], JUAN JOSÉ PIQUERAS[1], RAÚL PÉREZ[1], GREGORY BURWELL[1], IVAN NIKITSKIY[1],

TANIA LASANTA[1], TERESA GALÁN[1], ERIC PUMA[1],

ALBA CENTENO[3], AMAIA PESQUERA[3], AMAIA ZURUTUZA[3],

GERASIMOS KONSTANTATOS[1,2,*], FRANK KOPPENS[1,2,*]

**AFFILIATIONS:**

[1] ICFO-Institut de Ciencies Fotoniques, The Barcelona Institute of Science and Technology, 08860 Castelldefels (Barcelona), Spain.

[2] ICREA – Institució Catalana de Recerça i Estudis Avancats, Barcelona, Spain.

[3] Graphenea SA, 20018 Donostia-San Sebastian, Spain

*Correspondence to: gerasimos.konstantatos@icfo.eu , frank.koppens@icfo.eu

† These authors contributed equally to this work

**ABSTRACT:**

**Integrated circuits based on CMOS (complementary metal-oxide semiconductors) are at the heart of the technological revolution of the past 40 years, as these have enabled compact and low cost micro-electronic circuits and imaging systems. However, the diversification of this platform into applications other than microcircuits and visible light cameras has been impeded by the difficulty to combine other semiconductors than silicon with CMOS. Here, we show for the first time the monolithic integration of a CMOS integrated circuit with graphene, operating as a high mobility phototransistor. We demonstrate a high-resolution image sensor and operate it as a digital camera that is sensitive to UV, visible and infrared light (300 – 2000 nm). The demonstrated graphene-CMOS integration is pivotal for incorporating 2d materials into the next generation microelectronics, sensor arrays, low-power integrated photonics and CMOS imaging systems covering visible, infrared and even terahertz frequencies.**

## INTRODUCTION

The immense impact of microelectronics on our society is accredited to the miniaturization of silicon integrated circuits[1,2]. Alongside faster CPUs and higher capacity memories, it has enabled low-cost and high-performance digital imaging[3], with stunning pixel densities above 100 megapixels per chip[4,5]. More recently, the integration of photonics with CMOS electronic circuits is paving the way to large data communications bandwidths, higher connection capacities, and on-chip optical interconnects[6].

However, the difficulty in integrating non-silicon electro-optical materials with silicon integrated circuits has been a serious impediment to unlock its vast potential for imaging beyond the visible range, on-chip low-power optical data communications and compact sensing systems. Graphene and related 2d materials have shown their merits for a wide range of optoelectronic applications[7], such as data communications[8–10], high-performance LEDs[11], ultra-fast optical modulation[12] and photodetection with speeds up to 80GHz[13] and extreme broadband photodetection[14–21], for UV, visible, infrared and terahertz. Also, several high performance electronic devices and sensors have been demonstrated, such as ultra-sensitive Hall sensors[22], radio frequency receivers[23], strain sensors[24], biosensors[25], gas sensors[26], and high frequency transistors[27,28]. However, for these lab demonstrations, separate instruments performed the read-out. In order to unlock the true potential for these application areas, monolithic integration of 2d materials with CMOS integrated circuits is required as it enables compact and low-cost devices.

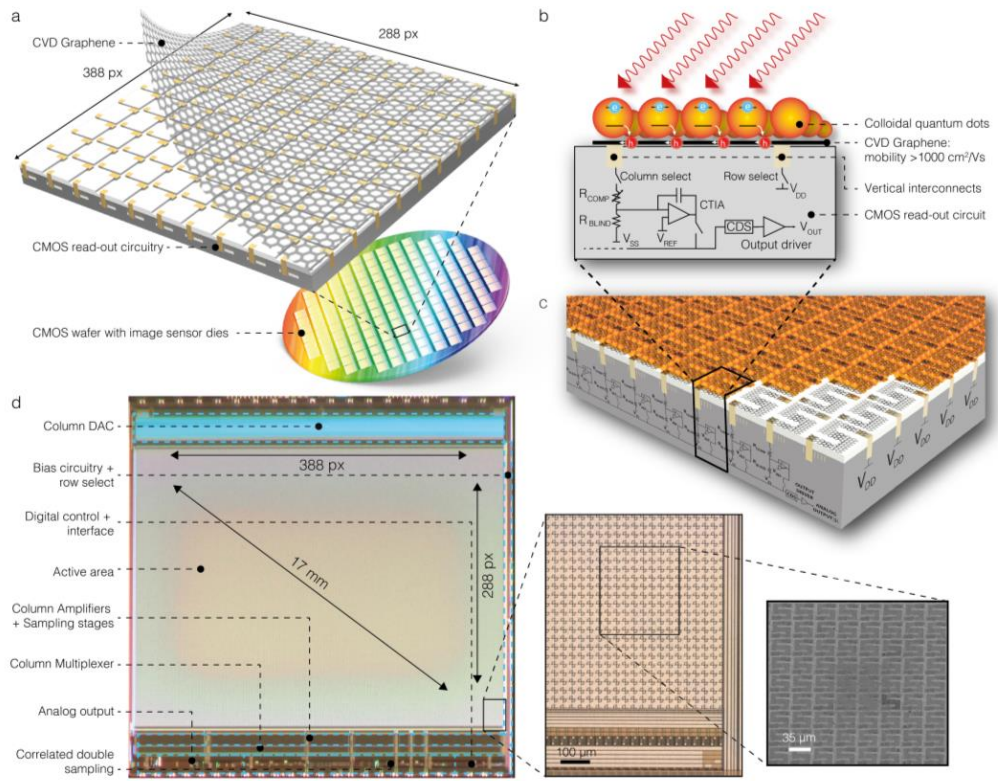

**Fig. 1 Back-end-of-line (BEOL) CMOS integration of CVD graphene with 388 x 288 pixel image sensor read-out circuit. a)** 3d impression of the CVD graphene transfer process on a single die containing an image sensor read-out circuit that consists of 388 x 288 pixels. **b)** Side view explaining the graphene photoconductor and the underlying read-out circuit. The graphene channels are sensitized to UV, visible, near infrared and short wave infrared light with PbS colloidal quantum dots: upon light absorption an electron-hole pair is generated, due to the built in electric field the hole transfers to the graphene while the electron remains trapped in the quantum dots. The schematic represents the CTIA based balanced read-out scheme per column and global correlated double sampling stage and output driver. **c)** 3d impression of the monolithic image sensor displaying the top level with graphene carved into s-shaped channels sensitized with a layer of quantum dots, vertical interconnects and underlying CMOS read-out circuitry. **d)** Photograph of the image sensor indicating the functionality for each area. To enhance contrast for different regions the photograph was taken before the colloidal quantum dots were deposited. Inset 1: Microscope image of the lower right corner of the active area of the ROIC. Inset 2: scanning electron micrograph of the active area of the image sensor displaying the s-shaped graphene channels. Both images were taken before the quantum dots were deposited.

One of the key advantages of 2d materials is that they can be transferred to virtually any substrate, and is therefore a key enabler for monolithic integration with silicon integrated circuits based on CMOS. This permits strong benefit from the technological maturity and time/cost effectiveness of CMOS wafer-scale production processes and the rapid developments of wafer-scale chemical vapour deposition (CVD) growth and transfer of graphene[23,29–31].

Here we present the first monolithic integration of graphene with a CMOS integrated circuit. In this case, the integration potential is shown by the realization of an image sensor with a 388x288 array of graphene-quantum dot photodetectors that is operated as a digital camera with high sensitivity for both visible and short-wave infrared light. The ~110,000 photoconductive graphene channels, are all individually integrated vertically, connecting to the individual electronic components of a CMOS readout integrated circuit (ROIC). The chip containing the circuitry is similar to those used for commercial image sensors in digital cameras[3], commonly used in smartphones, but here operating for both visible and short-wave infrared light (300 – 2000 nm). This wavelength range is so far not attainable with monolithic CMOS image sensors. Therefore, a broadband sensing platform that is monolithically integrable with CMOS is highly desirable. This proof-of-principle monolithic CMOS image sensor is a milestone for low-cost and high-resolution broadband and hyperspectral imaging systems[32], with applications in safety and security, smartphone cameras, night vision, automotive sensor systems, food and pharmaceutical inspection, and environmental monitoring[33].

**DEVICE STRUCTURE AND FUNCTIONALITY**

The integration of our CMOS graphene-quantum dot image sensor is a back-end-of-line (BEOL) process schematically shown in Fig. 1a. The process starts with a graphene transfer on a CMOS die that contains the read-out circuitry of the image sensor. Now, each pixel structure is covered with a layer of graphene that is connected with the bottom readout circuitry through vertical metal interconnects (Fig. 1b and c). Next, graphene is patterned to define the pixel shape as is shown in Fig. 1d, inset 2 and Supplementary Figure S1. Finally, a sensitizing layer of PbS colloidal quantum dots is deposited via a simple spin-casting process, on top of the graphene layer. The photoresponse is based on a photogating effect as follows[16,17] light is absorbed in the quantum dot layer followed by transfer of photo-generated holes (or electrons) to the graphene, where these circulate due to a bias voltage applied between the two pixel contacts (illustrated in Fig. 1b). Therefore, the photo-signal is sensed as a change in the conductance of the graphene transport layer. Due to the high mobility of graphene (here ~1000 cm$^2$/Vs), this photoconductor structure exhibits ultra-high gain of $10^8$ and responsivity above $10^7$ A/W, which is a strong improvement compared to photodetectors and imaging systems based on quantum dots only[34]. Our individual photodetector prototypes show detectivity above $10^{12}$ cm·√Hz/W (Jones) and spectral sensitivity from 300-2000 nm, and together with recently demonstrated switching times of 0.1-1ms clearly validate the applicability for infrared imaging[35]. Apart from the array of photosensitive pixels, the imager contains a row of blind pixels that are used to subtract the dark signal as the photodetectors are voltage biased. We remark that here, the spectral range is determined by the quantum dot material and size, but this approach can be generalized to other types of sensitizing materials in order to extend or tune the spectral range of the sensor element.

The functional elements of the CMOS circuitry are shown in Fig 1b, c and d. The elements surrounding the active pixel area provide multiple functions: signal path control, photodetector biasing, tuneable compensation resistors, blind pixels, amplification and read-out of the photo-signal from pixel to output, and control of the image exposure and shutter operation. The photosignal per pixel is acquired through a balanced read-out scheme as shown in the schematic in Fig 1b, that consists of the blind pixel (with resistance $R_{blind}$) and a tuneable compensation resistance $R_{comp}$ in series with the pixel resistance $R_{pixel}$ that can be digitally controlled for each individual pixel. Pixels are addressed sequentially on a row-by-row basis (rolling shutter) with a frame rate of maximally 50 frames per second (fps). The signal readout chain (see Fig. 1b and Supplementary Figure S5) is based on a Capacitive Trans-Impedance Amplifier (CTIA) per column that integrates the current difference between photosensitive and blind pixels. The amplifier output is sampled, before and after exposure, in a storage block, also per column, and all column signals are multiplexed into a common output bus terminal. Finally, a correlated double sampling (CDS) correction is performed to reduce readout noise and the resulting output signal $V_{out}$ is sent to the imager's analogue output.

**DIGITAL CAMERA**

We first present the main results of our work in Fig. 2, which encompass several types of images that have been captured with our prototype digital camera comprising the graphene-CMOS image sensor. The configuration for obtaining these images is schematically illustrated in Fig. 2a: the graphene-QD image sensor captures reflection images from objects illuminated by a light source of visible or short-wave infrared light. The grey scale plots of Fig. 2 are compiled of the normalized photosignals for each of the photodetection pixels of the 388 x 288 array, amplified and multiplexed by the CMOS integrated circuit. Not the entire active area of the image sensor is covered with graphene due to the finite size of the CVD graphene sheet and manual alignment of the transfer (see Supplementary Notes and Figure S7); the pixels that were not covered with graphene and hence did not show any conductance are represented as continuous grey areas. The image shown in Fig. 2c has been obtained using an image sensor with quantum dots that have an exciton peak at 920 nm, corresponding to the peak absorption of the quantum dots as measured in solution. The objects were illuminated with visible light with illumination power of ~$1 \cdot 10^{-4}$ W/cm$^2$, which corresponds to office illumination conditions. We remark that a reasonable fraction of the pixels was sensitive to much lower light levels (further discussed below), but the pixel drift and spread in sensitivity were too large to obtain extreme low-light level images. Further optimization of the fabrication process and wafer-scale processing can resolve these non-uniformities. The images shown in Fig 2 b,d,e,g,i have been obtained using an image sensor with CQDs that have an exciton peak at 1670 nm. For the image in Fig 2 b,e,g,i we illuminated the objects with an incandescent light source and filtered all the visible light from <1100 nm (Supplementary Methods). For the image in Fig 2d we used the full spectrum of the incandescent source to illuminate the scene to demonstrate the capability to capture VIS, NIR and SWIR light with one camera. In Fig 2e,g,i we show different use cases of a SWIR camera: vision under difficult weather circumstances (Fig 2e), silicon CMOS wafer inspection (Fig 2g) and water detection for food inspection (Fig 2i). The capability to capture short-wave infrared images demonstrates imaging applications that are impossible with the widely used silicon CMOS cameras.

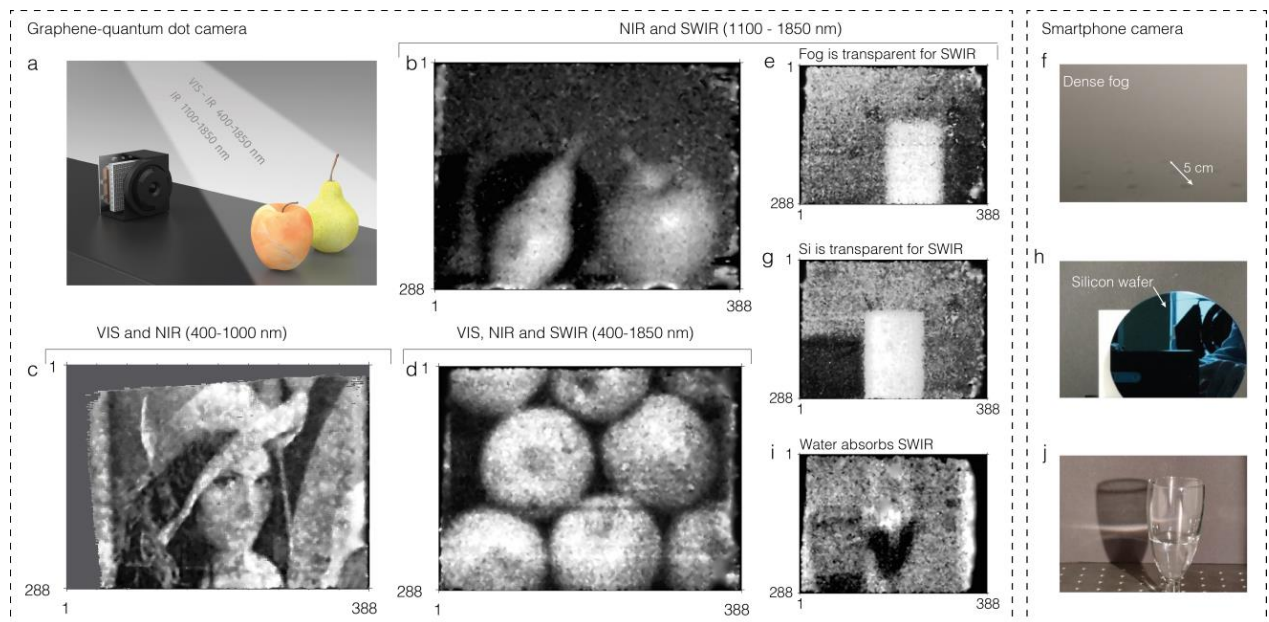

**Fig. 2 Hybrid graphene and colloidal quantum dot based image sensor and digital camera system a)** Digital camera setup representation: the image sensor plus lens module captures the light reflected off objects that are illuminated by an external light source. Supplementary Figure S3 contains all the details of the image capturing setup for each of the images in this figure. **b)** Near infrared (NIR) and short wave infrared (SWIR) light photograph of an apple and pear. An incandescent light source (1000W, 3200K) illuminated the objects. As this image sensor is sensitive to visible (VIS), NIR and SWIR light (300-1850 nm, Figure 4b) we placed an 1100 nm long pass filter in the optical path to reject all light that a conventional Si-CMOS sensor can capture. The axis tick labels indicate the column (horizontal axis) and row (vertical axis) numbers. The illumination yielded an irradiance on the image sensor of $\sim 1 \cdot 10^{-4}$ W/cm$^2$. The grey scale represents the photosignal dV in volts (dV=$V_{out,light}$- $V_{out,dark}$, Supplementary Methods) normalized to dV obtained from a white reference image. An image processing scheme as described in the Supplementary Methods has been performed. **c)** VIS, NIR and SWIR photograph of a box of apples, illuminated with the same source as in b), but without the 1100nm long pass filter. **d)** VIS and NIR (this image sensor is sensitive to 300-1000 nm, Figure 4a) photograph of standard image reference 'Lena' printed in black and white on paper illuminated with an LED desk lamp. **e)** NIR and SWIR image of a rectangular block covered in fog as shown in **f)**. The same source as in a) illuminated the scene. **g)** NIR and SWIR image of a rectangular block behind a silicon wafer as shown in **h)**. The same source as in a) illuminated the scene. **i)** NIR and SWIR image of a glass of water as shown in **j)**. The same source as in a) illuminated the scene. A smartphone camera captured images f), h) and j) under similar lighting conditions..

## MONOLITHIC GRAPHENE CMOS INTEGRATION PERFORMANCE

Key to graphene integration with the CMOS electronic circuit is a reliable and high-quality electrical connection between all (~111,744) the graphene conduction channels and the integrated circuitry trough vertical metal plugs. A 2d map of the conducting pixels and non-conducting pixels, obtained through the read-out circuitry of the ROIC, is shown in Fig 3a. Within the area where graphene is present, a resistance was recorded for 99.8% of the pixels, and hence the yield of the transfer, channel patterning and contacting is close to unity. In addition, it is important to match the resistance of the photosensitive pixels with the blind pixels, in order to properly subtract the dark signal. To this end, we have chosen a S-shape graphene channel targeting a resistance of 20kΩ to make it compatible with this specific ROIC that has an operation regime from 20-100 kΩ. (We remark that the S-shape graphene channel limits the fill factor of the photoactive area. This limitation can be solved by proper ROIC design optimized for graphene-QD photodetectors, in principle allowing for fill-factors close to 100% (see Supplementary Notes and Figure S9).) Using that same read-out circuitry, we could also obtain the resistance values for each of the conducting pixels. The inset of Fig. 2a shows that the resistance values in a 10x10 pixel area are all close to 20 kΩ. The variation in the pixel resistance ($R_{pixel}$) is most likely limited by the shortcomings of the wet transfer technique such as unwanted strain effects and unintentional doping[36].

The issue of variable graphene device resistance is ameliorated by utilizing the tuneable series resistor ($R_{comp}$) in the CMOS circuit, which can be digitally addressed and optimized for each individual pixel (Supplementary Notes and Figure S6). The results of this tuneable matching are visible in the resistance histograms shown in Figure 3b. This histogram reveals that after optimisation with the compensation resistor $R_{comp}$ for each pixel, a rather narrow distribution of pixel resistances is obtained with a spread of ~ 4 kΩ around an average value of 20 kΩ. This moves the resistance of most of the pixels inside the operation regime of the ROIC. Moreover, a great improvement in the detector yield, performance and uniformity across all the pixels is achieved.

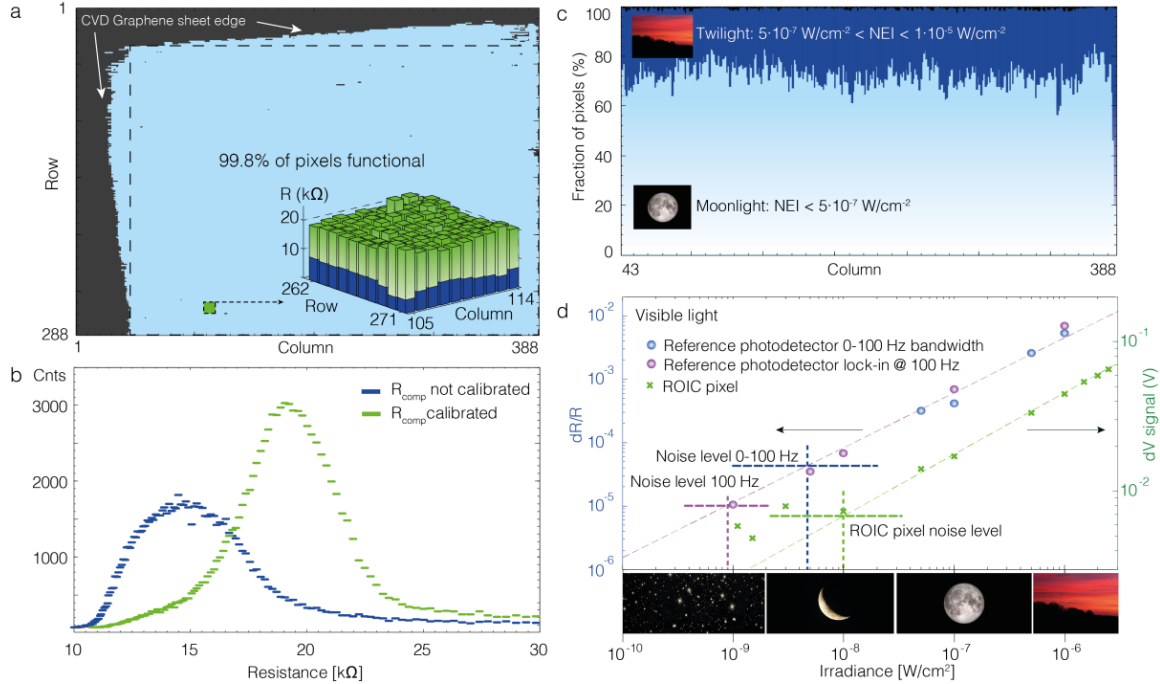

**Fig. 3 Electro-optical characterization a)** Map of the conducting pixels (light blue) and non-conducting pixels (grey). The box indicated with a dashed line indicates the area over which the yield was calculated. Inset: 2d barplot of the total resistance per pixel ($R_{pixel}$ + $R_{comp}$) values for a 10x10 pixel area (green square). Green: $R_{pixel}$, blue: $R_{comp}$. **b)** Histogram of $R_{pixel}$ before resistance compensation in blue and after compensation ($R_{pixel}$ + $R_{comp}$) in green. $R_{comp}$ varies from 0 to 8 kΩ. **c)** Histogram of NEI for all pixels inside the dashed box in figure 3a, plotted per column (in total 255 pixels for each column). Light blue: pixels that are sensitive to moonlight, dark blue: pixels that are sensitive to twilight, black: pixels that are not sensitive to light. **d)** Photoresponse versus power at uniform illumination with λ = 633 nm and measured from twilight (~$10^{-6}$ W/cm²) down to starlight ($10^{-10}$ – $10^{-9}$ W/cm²) conditions. The green crosses are data obtained from a representative pixel in the image sensor. Blue and purple circles represent photoresponse (expressed in light-induced resistance change dR/R) of a reference photodetector with the same type of colloidal quantum dots and a channel of 48 μm width and 8 μm length. The datapoints in blue are obtained using a DC-coupled amplifier with a bandwidth of 100 Hz. The data in purple are obtained using a lock-in type measurement technique at 100 Hz modulation. The images below the plot illustrate the illumination conditions: from star light to moon light to twilight.

This can be inferred from detector performance measurements, which monitor the dependence of the detector signal on the irradiance for each pixel. In combination with measurements of the detector noise (Supplementary Figure S8), the key figure of merit for the detector sensitivity can be extracted: the noise-equivalent irradiance (NEI). An example for one specific pixel is shown in Fig. 3d, from which we infer a photoresponse down to a NEI level of $10^{-8}$ W/cm², which corresponds to the irradiance from a quarter moon.

A complete map of the sensitivities (expressed in terms of NEI) of all the pixels of the imager is shown in Fig. 3c, revealing that more than 95% of the pixels is sensitive to irradiance corresponding to partial moon and twilight conditions in the visible range (for a wavelength of 633 nm). The dynamic range of the imager is limited by the readout circuit because the graphene-quantum dot pixel conductance does not match the optimum point for which this off-the-shelf ROIC has been designed. For comparison, we show the complete electro-optical characterization of the image sensor as well as individual reference photodetectors in Fig. 3d, 4d and Supplementary Table S1. We find that low-frequency 1/f noise dominates the detector noise (Supplementary Figure S8), but owing to the high responsivity, high sensitivities (and thus low NEI) have been obtained. Reference photodetectors can operate stable at much smaller NEI (down to $10^{-9}$ W/cm$^2$), have a dynamic range above 80 dB and speed above 100 fps (inset Fig 4d). Customizing the readout circuitry would allow reaching the sensitivity and dynamic range of an individual reference photodetector in an imaging system. Therefore, customized readout chips would enable operation at higher frame rates, and detection sensitivities that are comparable to commercial imaging systems, but with the main advantage that our system is CMOS-monolithic and sensitive to UV, visible, near infrared and short-wave infrared light.

**BROADBAND IMAGING SYSTEM**

We further demonstrate this extensive spectral range in Figs. 4a,b, showing the photoresponse spectra for two different imagers based on two different sizes of quantum dots that are deposited on top of the graphene layer. The spectra show exciton peaks at 920 nm and 1670 nm. The detection sensitivity reaches from UV up to 1850 nm. To demonstrate the capability for night vision applications, Fig. 4c shows the detection by an individual detector of the "night glow", which is the emission of short wave infrared light by the earth's atmosphere that can be utilized for passive night vision[37]. By aiming our detector at a dark and clear sky, the night-glow signal in Fig. 4c is clearly distinguished from the noise (Supplementary Methods and Figure S4).

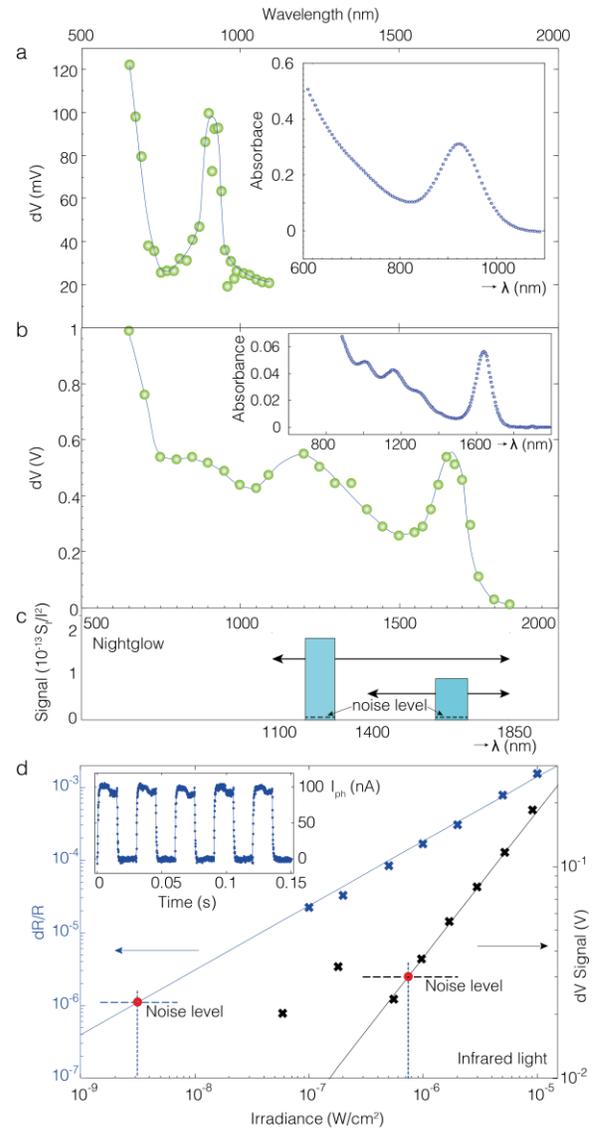

Fig. 4 Visible (VIS) and short wave infrared (SWIR) sensitivity and night glow measurement. a) Spectral dependence of the photoresponse for one of the pixels of the ROIC sensitized with quantum dots that have an exciton peak at 920 nm measured at a constant irradiance of 5·10$^{-5}$ W/cm$^2$. The continuous blue line is a guide to the eye. The inset shows the absorbance spectrum of the quantum dots in solution. b) Spectral dependence of the photoresponse for one of the pixels of the ROIC sensitized with quantum dots that have an exciton peak at 1670 nm measured at a constant irradiance of 10$^{-4}$ W/cm$^2$. The inset shows the absorbance spectrum of the quantum dots in solution. c) Measurement of the nightglow using a SWIR sensitive reference photodetector of 1x1 mm$^2$, aiming at a dark, clear sky for long-pass filtering with a cut-off at 1100 nm and 1400 nm (see SOM). The dashed lines indicate the noise level obtained with a lock-in measurement modulated at 10 Hz. d) Photoresponse versus power for a reference photodetector illuminated with 1550 nm light (blue crosses, left vertical axis) and for the ROIC pixels in the SWIR regime, illuminated with 1670 nm light (black crosses, right vertical axis). The datapoints in blue are obtained using a DC-coupled amplifier with a bandwidth of 35 Hz. The reference photodetector exhibits the exciton peak at 1580 nm and has a channel of 1x1 mm$^2$. Inset: photocurrent versus time of the reference photodetector illuminated with 1550 nm light at an irradiance of 3·10$^{-5}$ W/cm$^2$, sampled at 10 kS/S.

**CONCLUSIONS AND OUTLOOK**

Future graphene-based image sensors can be designed to operate at higher resolution, in a broader wavelength range, and potentially even with a form factor that fits inside a smartphone or smartwatch (Supplementary Notes, Figure S9). In contrast to current hybrid imaging technologies (which are not monolithic), we do not encounter fundamental limits with respect to shrinking the pixel size and increasing the imager resolution. Graphene patterning and contacting, i.e. lithography, will ultimately be the limiting factor. Therefore, competitively performing image sensors with multi-megapixel resolutions and pixel pitches down to 1 μm are within reach (Supplementary Table S2).

The confirmation that a complex graphene - CMOS circuit is operational paves the way for a wide range of electronic and opto-electronic technologies where monolithic integration is essential, such as integrated photonics, high frequency electronics, and arrays of sensors. Future development of the transfer and encapsulation of graphene (for example, based on hBN[36]) will further increase the uniformity and performance of graphene-CMOS technologies. A further compelling future prospect is the creation of 3d integrated circuits based on 2d materials with Si-CMOS that can perform even more complex tasks. For example, the natural layer-by-layer stackability of graphene and other 2d materials opens a wealth of possibilities to add electronic and opto-electronic functions in the vertical dimension; all integrated into CMOS microelectronic and optoelectronic circuits.